\documentclass[english]{emulateapj}
\usepackage{babel}

\usepackage{hyperref}
\usepackage{amsmath}
\usepackage{amssymb}

\newcommand{\sn}[2]{\ensuremath{#1 \times 10^{#2}}}
\def\eps@scaling{.95}

\begin{document}

\title{Changing Ionization Conditions in SDSS Galaxies with AGN as a Function of Environment from Pairs to Clusters}

\author{Emil T. Khabiboulline$^{\ast,1,2}$, Charles L. Steinhardt$^{1,2}$, John D. Silverman$^2$, Sara L. Ellison$^3$, J. Trevor Mendel$^4$, and David R. Patton$^5$}
\affil{$^\ast$ekhabibo@caltech.edu $^1$California Institute of Technology $^2$Kavli Institute for the Physics and Mathematics of the Universe $^3$Department of Physics and Astronomy, University of Victoria $^4$Max Planck Institute for Extraterrestrial Physics $^5$Department of Physics and Astronomy, Trent University}

\begin{abstract}

We study how AGN activity changes across environments from galaxy pairs to clusters using $143\, 843$ galaxies with $z<0.2$ from the Sloan Digital Sky Survey (SDSS). Using a refined technique, we apply a continuous measure of AGN activity, characteristic of the ionization state of the narrow-line emitting gas. Changes in key emission-line ratios ([N{\small II}]$\lambda6548$/H{\small $\alpha$}, [O\small{III}]$\lambda5007$/H{\small $\beta$}) between different samples allow us to disentangle different environmental effects while removing contamination. We confirm that galaxy interactions enhance AGN activity. However, conditions in the central regions of clusters are inhospitable for AGN activity even if galaxies are in pairs. These results can be explained through models of gas dynamics in which pair interactions stimulate the transfer of gas to the nucleus and clusters suppress gas availability for accretion onto the central black hole.

\end{abstract}

\section{Introduction}
\label{sec:Introduction}

Supermassive black holes grow through the accretion of matter, predominantly gas~\citep{Lynden-Bell1969}. This accretion powers active galactic nuclei (AGN), the highly luminous centers of certain galaxies. The total accretion of matter inferred from the AGN luminosity throughout cosmic history is approximately equivalent to the black hole mass density in the local universe~\citep{Soltan1982}, implying that analyzing luminous accretion equates to studying black hole growth. Moreover, AGN produce extraordinarily salient signals through light emission that is indicative of the nature of accretion, leading to classifications such as quasars, radio galaxies, Seyferts, and LINERs (although not all LINERs are AGN~\citep{Yan2013,Singh2013}). Studies indicate correlations between the state of the AGN and the properties of the host galaxy, thus intrinsically making AGN key to understanding the black hole-galaxy connection.

It has been well established that environmental processes influence the stellar mass growth of galaxies~\citep{Peng2010}. Star formation is induced by close encounters, such as mergers, between galaxies~\citep{Woods2007,Kampczyk2013}. Whether such external factors can also impact nuclear activity by determining how much gas is around the black hole (availability) and how much falls in (delivery) is a key open question that has begun to be addressed. 

Simulations of black hole growth show that mergers trigger AGN~\citep{DiMatteo2005,Foreman2009}. \citet{Hernquist1989,Domingue2005} carried out simulations that suggest higher rates of gas inflow due to merging as the cause. In addition, observations show that pair interactions of galaxies correspond to increased AGN activity. \citet{Silverman2011}, using the zCOSMOS survey, \citet{Ellison2011}, using the Sloan Digital Sky Survey (SDSS), and \citet{Koss2010}, using the Swift BAT survey, demonstrate that the likelihood of AGN is higher in nearby pairs of galaxies, many of which are about to merge, as opposed to isolated ones. \citet{Ellison2013} report highest black hole accretion rates in the closest pairs and post-mergers, which \citet{Satyapal2014} supports with a mid-infrared study including obscured AGN. Identifying mergers from a pairs sample, \citet{Alonso2007,Cotini2013} also find increases in AGN fraction and accretion rate. Using a new method to identify mergers by the presence of two close nuclei before final coalescence, \citet{Lackner2014} show that such mergers boost both star formation and AGN activity by a similar factor. Furthermore, merger simulations are consistent with the observations of double quasars at small separations~\citep{Foreman2009}. While pair interactions have been shown to correlate with AGN activity, secular processes contribute significantly to the fueling of the supermassive black hole~\citep{Reichard2009,Draper2012,Kocevski2012,Schawinski2012}.

The rich environment of galaxy clusters may also influence AGN activity (in a different manner) but its impact is less clear. Over the redshift range $z=0.2$-$0.7$, \citet{Ruderman2005} find a spike in the number of AGN at the centers of clusters, attributed to close encounters between infalling galaxies and the large central cD-type elliptical galaxy, as well as a broad secondary excess around the virial radius, attributed to galaxy mergers. Meanwhile, \citet{Pimbblet2013} report that AGN fraction increases from the cluster center to 1.5 $R_{virial}$, tailing off at higher radii. This trend is attributed to a changing mix of galaxy types as a function of radius. An analysis of supercluster A109/2 at $z\sim 0.17$ agrees by finding that AGN galaxies lie mainly in environments comparable to cluster outskirts, with no AGN found in the areas of highest or lowest galaxy density~\citep{Gilmour2007}. However, other studies also using a measure of local galaxy density conclude that the AGN fraction is constant from the cluster center to the rarefied field~\citep{Miller2003,Sorrentino2006}. Several of these opposing results are discussed by \citet{Martini2007}, who demonstrate that the highest-luminosity AGN are more centrally concentrated than inactive galaxies but also that the effect disappears when analyzing a wider range of luminosity, and the AGN fraction is not lower in clusters compared to the field for X-ray and radio-selected AGN but is lower for optically-selected AGN.

Environmental effects associated with galaxy pairs and clusters are likely not independent but rather have a complicated interplay. Clusters, acting on a larger scale than pairs, might affect AGN activity by virtue of their effect on pair interactions. Additional physical factors may affect the gas content of galaxies that provides the fuel reservoir, including ram pressure stripping, tidal interactions, harassment, and strangulation. Therefore, a combined analysis of environmental factors is important. Measurements of environment density have been used to show that some types of AGN (e.g., radio-loud) tend to reside in over-dense environments~\citep{Karouzos2014} and that triggered star formation only occurs in relatively low density regions~\citep{Ellison2010}. Both optical~\citep[i.e., type 2; ][]{Kauffmann2004} and X-ray-selected~\citep{Silverman2009b} AGN show a preference for the low-density environment most pronounced for the massive galaxies ($M_{stellar}>10^{11} M_{\sun}$). \citet{Sabater2013} performed a combined analysis, finding that AGN fraction increases due to pair interactions and decreases in denser environments, as in clusters.

Previous studies, including those described above, have generally grouped similar galaxies together and analyzed fractions based on a discrete classification of galaxies into broad categories (e.g., fraction of AGN galaxies, fraction of star-forming galaxies). While this method has been shown to work given sufficient statistics, a continuous metric may instead be a better description of the impact of environment on AGN strength, as not all star-forming galaxies or all AGN are identical.  Such a metric will enable a full exploitation of the available information, meaning that the same sample can provide results with greater statistical significance and the discovery of small effects that were previously impossible to detect. \citet{Ellison2013} used a continuous quantity, L[O{\small III}], to find that AGN accretion rate increases towards smaller pair separations, peaking in post-mergers. In this study, we develop and employ a continuous metric of AGN activity, which is sensitive to the ionization conditions of the interstellar medium of galaxies hosting AGN and likely tracks the AGN strength, and thus accretion rate. By studying how this activity varies across the AGN population in pairs and clusters, we hope to form a more general picture of how the intergalactic environment influences black hole accretion in AGN. 

We carry out a joint analysis of the relationship between close galaxy pairs, the cluster environment, and AGN activity using the SDSS DR7 spectroscopic survey. In \S~\ref{sec:Dataset}, we describe the dataset of galaxies that we analyze. Then, \S~\ref{sec:AGNMeasure} describes how we continuously measure variation in AGN activity among these galaxies due to environmental processes. We present the results of a series of studies in \S~\ref{sec:Results}, showing that pairs and clusters exhibit a small increase and a large decrease in AGN activity, respectively, with the clusters' influence being dominant in the central regions. Finally, in \S~\ref{sec:Discussion}, we discuss how these conclusions might be explained by gas dynamics assuming that close pair interactions increase delivery of gas and a cluster environment decreases availability of gas for accretion onto the central black hole.

\section{Data}
\label{sec:Dataset}

The data used in this work are drawn from a compilation of catalogs based upon the Sloan Digital Sky Survey (SDSS). The SDSS Data Release 7 (DR7) catalog includes $930\, 000$ galaxies~\citep{Abazajian2009}, with additional information on these galaxies provided by ancillary catalogs (e.g., from the Max-Planck Institut f\"{u}r Astrophysik and Johns Hopkins University (MPA-JHU) collaboration\footnote{http://www.mpa-garching.mpg.de/SDSS}). We use a sample of galaxy pairs with separations less than 80 kpc selected from the SDSS that has been previously used to investigate many properties of merging galaxies~\citep[e.g.,][]{Ellison2008,Ellison2010,Ellison2011,Patton2011,Scudder2012,Satyapal2014}. Clustering information comes from the \citet{Yang2007} catalog, updated for SDSS DR7. This compilation provides a large dataset with substantial environmental information, something that has only recently become possible.

For all of these galaxies, stellar masses are derived from the improved photometry of \citet{Simard2011}, and the observed spectral energy distribution is compared to a library of synthetic stellar populations~\citep{Mendel2014}. We set a cut on redshift (SDSS-measured) at $z < 0.2$; for greater values the sample is increasingly incomplete~\citep{Patton2008}. The emission line fluxes used ([N{\small II}], [O{\small III}], H{\small $\alpha$}, and H{\small $\beta$}) are the MPA-JHU values~\citep{Brinchmann2004}, corrected for Galactic extinction, internal extinction (using the SMC extinction curve), and continuum absorption. In addition, we perform a weak quality cut on the emission lines by requiring a signal-to-noise ratio of 1 or greater. \citet{Kauffmann2003} separate galactic spectra dominated by AGN activity (AGN galaxies) from galaxies dominated by star formation (SF galaxies) using the Baldwin, Phillips \& Terlevich (BPT)~\citep{Baldwin1981} diagnostic diagram. We limit our analysis of AGN activity only to the AGN-classified galaxies (cf. \S~\ref{subsec:D} for details).

Galaxy pairs are identified using the techniques in \citet{Ellison2011}. For a galaxy to be classified as a pair galaxy, we impose upper limits on separation (80 $h_{70}^{-1}$ kpc), stellar mass ratio (4:1), and line-of-sight velocity difference (300 km s$^{-1}$) for the identified pair. To account for spectroscopic incompleteness at separations $<55$ arcsec due to fibre collisions, a random $67.5\%$ of pairs with separations $>55$ arcsec are excluded, following \citet{Ellison2008}. Nonpair galaxies are defined as those galaxies that do not have a close companion within 80 $h_{70}^{-1}$ kpc and $10\, 000$ km s$^{-1}$. In addition, the Galaxy Zoo~\citep{Lintott2008,Lintott2011} merger vote fraction must be 0~\citep{Darg2010,Ellison2013b} for our nonpair sample, meaning that, upon visual inspection of the galaxies, no citizen scientists classified them as possible mergers. We note that while nonpair galaxies are definitely not merging, pair galaxies are not necessarily merging galaxies. Indeed, $47\%$ of our pair galaxies have a merger vote fraction of 0. Hence, ours is not a direct study of merging, but rather of the effects of all close pair interactions, including merging.

Cluster determination is based on the \citet{Yang2007} catalog. The group-finding algorithm~\citep{Yang2005} classifies groups as galaxies that reside in the same dark matter halo, using an enhanced Friends-of-Friends algorithm that takes into account galaxy kinematics. There is no established definition for a cluster, however, since all galaxies are put into groups, even if the group only has one member. Richness, determined as the number of members in a group, is used to classify groups into clusters and nonclusters. Galaxies in groups with a richness of 10 or greater are classified as cluster galaxies, those in groups with a richness of 1 are classified as noncluster galaxies, and the rest with intermediate richness are excluded from the analysis due to the ambiguity in their true nature. 

Using richness this way gives conservative cluster selection. Any dependence of richness on redshift is addressed through the matching procedure described in \S~\ref{subsec:DeltaD}. Using group halo mass instead of richness leads to classification errors. For example, a single massive galaxy could be classified as a large group due its large associated halo mass. Moreover, groups with only a few galaxies, which we would not normally consider to be clusters, are the overwhelming majority and thus strongly dominant over the true clusters over the entire halo mass range, while with increasing richness groups quickly develop greater halo mass. Richness also likely is more sensitive to local pair interactions within the cluster, which are more probable for larger numbers of grouped galaxies, rather than a cluster-wide effect dependent on total halo mass, which could be dominated by just one or two very massive galaxies. While we use richness to conservatively classify groups into clusters, further classification of clusters into relative sizes is done using both richness and halo mass (cf. \ref{subsec:Clusters}), since for this conservative subsample either measure accurately gives size.
 
The final sample consists of $3\, 151$ pair galaxies, $108\, 700$ nonpair galaxies, $9\, 530$ cluster galaxies, and $101\, 824$ noncluster galaxies. There is redundancy (i.e., most galaxies fall into the two categories of pair/nonpair and cluster/noncluster) but also some incompleteness (some pair/nonpair galaxies are not classified as cluster/noncluster). Overall, there are $143\, 843$ distinct galaxies.

\section{A Continuous Measure of Change in AGN Activity}
\label{sec:AGNMeasure}

Because this study focuses on how an AGN galaxy's environment affects its activity, establishing a suitable way to measure and compare the strength of AGN across environments is vital. Measuring AGN activity in general is a difficult problem in observational astronomy, particularly as the diagnostics are often limited to those within the optical spectrum. Many efforts have been made to pick out AGN and their activity. Depending upon how much of the spectrum is available, the equivalent width of the [O{\small III}] emission line and classification by the BPT diagnostic diagram, which relies on the [N{\small II}]/H{\small $\alpha$} and [O{\small III}]/H{\small $\beta$} line ratios, are commonly used. These are well-established, with [O{\small III}] equivalent width a measure of a mixture of star formation and AGN activity \citep{Kauffmann2003} and BPT classification employed to identify the relative contributions of star formation and AGN activity. In its traditional form, the BPT diagram is used as a classifier rather than providing a continuous measure of AGN activity.

To make maximal use of this information, we seek a measure that isolates AGN activity and indicates its strength, rather than merely its presence.  The current use of a Boolean classification loses substantial information on the strength of the AGN, which could have been extracted from the continuous spectral data. A continuous measure also enables us to use powerful statistical tests in order to find correlations and determine their statistical significance.  Ultimately, in this work we will use this new measure to indicate the continuous relationship between the level of AGN activity changes and the environment of the host. 

\subsection{\texorpdfstring{$D$ Parameter}{D Parameter}}
\label{subsec:D}

One of the most promising continuous measures of AGN activity in the literature is based on the BPT diagnostic diagram. The BPT diagram plots galaxies on the [O{\small III}]/H{\small $\beta$} and [N{\small II}]/H{\small $\alpha$} coordinate axes in order to essentially form a continuous distribution from SF to AGN galaxies. The two types are generally mapped onto two wings on the diagram, with AGN galaxies on the right, but the precise boundary is not clear. \citet{Kauffmann2003} empirically construct a classification curve (Eq.~\ref{eq:K03}) from a complete SDSS sample. Earlier, \citet{Kewley2001} used photoionization models to capture the position of even the most extreme SF galaxies (Eq.~\ref{eq:K01}). Galaxies found in between the two curves are referred to as ``composites,'' since they may have contributions from both star formation and nuclear activity. ``Pure'' AGN are those above the \citet{Kewley2001} curve.

\footnotesize
\begin{equation}
\log(\text{[O{\small III}]}/\text{H{\small $\beta$}})=0.61/(\log(\text{[N{\small II}]}/\text{H{\small $\alpha$}})-0.05)+1.30
\label{eq:K03}
\end{equation}
\begin{equation} 
\log(\text{[O{\small III}]}/\text{H{\small $\beta$}})=0.61/(\log(\text{[N{\small II}]}/\text{H{\small $\alpha$}})-0.47)+1.19
\label{eq:K01}
\end{equation}
\normalsize

\begin{figure}[!hbtp]
  \centering
  \leavevmode
  \includegraphics[width={\eps@scaling\columnwidth}]{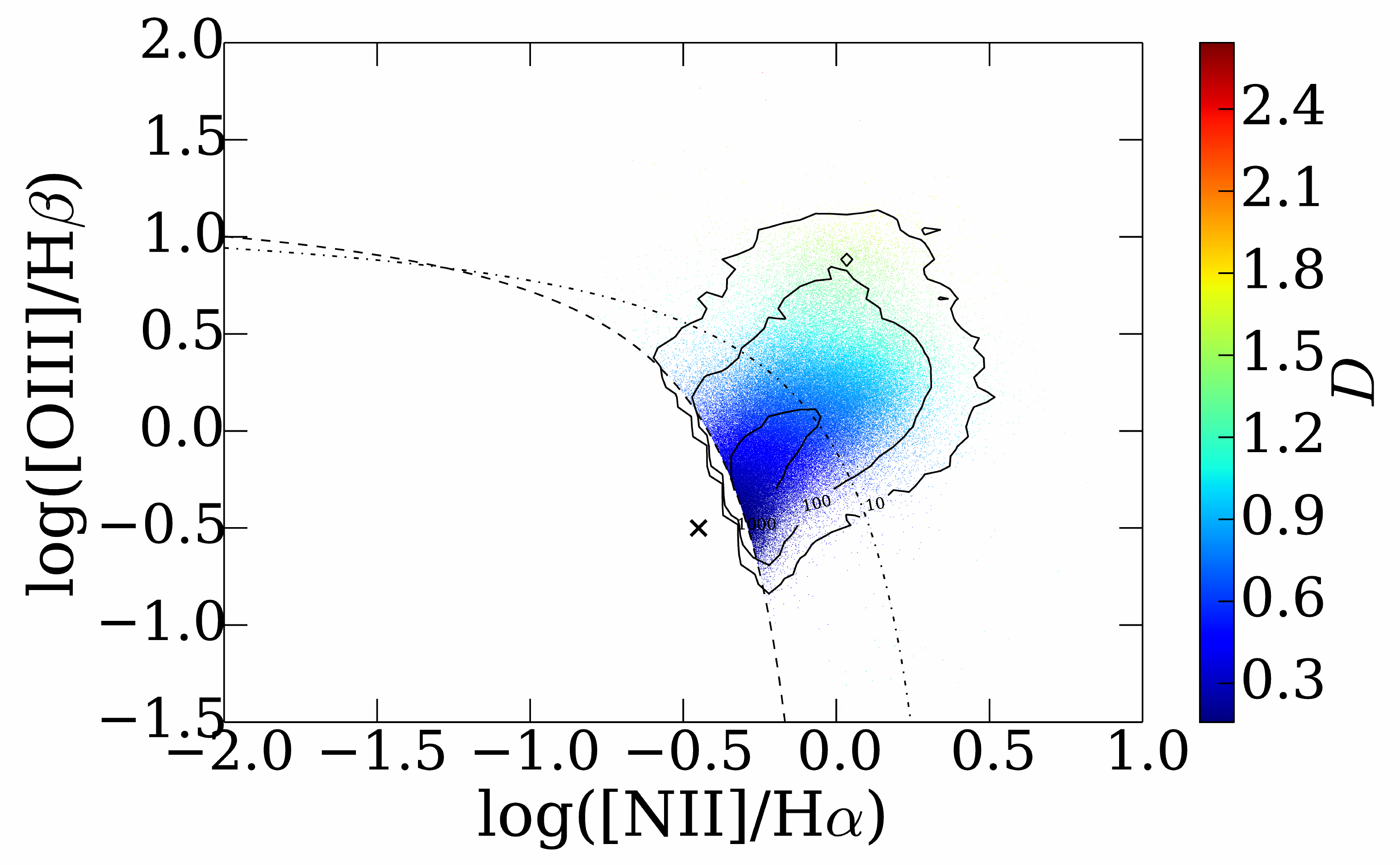}
  \caption{BPT diagram of all sample galaxies shown as individual points on the [O{\scriptsize III}]/H{\scriptsize $\beta$} and [N{\scriptsize II}]/H{\scriptsize $\alpha$} coordinate axes. Color indicates $D$ value, which is distance from a designated point deemed the SF locus (black x), at coordinates $(-0.45,-0.5)$. AGN galaxies lie above the \citet{Kauffmann2003} classification curve (dashed black curve). Those below the \citet{Kewley2001} classification curve (dash-dotted black curve) are composites that may have significant star formation activity, while those above are pure AGN galaxies where AGN activity is dominant. Density contours (solid black curves) correspond to a $70 \times 60$ rectangular grid (not shown) and indicate densities (10, 100, 1000 per rectangle) of galaxies on the diagram.}
  \label{fig:DOnBPT}
\end{figure}

As an example, we show our sample, consisting of composite and pure AGN galaxies, on the BPT diagram (Fig.~\ref{fig:DOnBPT}). With increasing distance from the SF locus, galaxies display increasingly AGN-like behavior. Working on this premise, the optical ``$D$'' parameter (Eq.~\ref{eq:D}) is the distance from the center of the locus of SF galaxies on a BPT diagram~\citep{Kauffmann2003}. Pure AGN galaxies have the highest $D$, as expected since their ionization field is dominated by AGN activity. Composite AGN galaxies have smaller $D$, reflecting a lower estimate of AGN activity in light of a higher contribution from star formation.

\footnotesize
\begin{equation} 
D=\sqrt{[\log(\text{[N{\scriptsize II}]}/\text{H{\scriptsize $\alpha$}}) + 0.45]^2 + [\log(\text{[O{\scriptsize III}]}/\text{H{\scriptsize $\beta$}}) + 0.5]^2} 
\label{eq:D}
\end{equation}
\normalsize

The BPT diagram works because different line ratios correspond to different combinations of spectral hardness and intensity, which are the signature of the incident radiation field. Because of the higher energies involved in accretion, objects with harder spectra are predominantly dominated by AGN and those that have more overall activity display more intense lines. The $D$ parameter correlates closely with [Ne{\small V}]/[Ne{\small II}], a probe of incident radiation field hardness and a signature of AGN, in addition to several other infrared diagnostics of AGN activity~\citep{LaMassa2012}. These infrared measures are independent of the optical as tests of the radiation field. Moreover, with increasing distance from the star-forming sequence, the hardness of the ionizing radiation field increases~\citep{Kewley2006}.  Thus, it is natural to hope that the $D$ parameter might provide a good indication of the level of AGN activity.

Given that the purpose of the BPT diagram is essentially one of classification, it might be argued that $D$ reflects the balance between AGN and star formation activity~\citep{Kauffmann2003} rather than directly measures AGN activity. Although studies show that $D$ correlates closely with AGN activity, and with radiation field hardness in particular, it may also correlate to some extent with star formation. In addition, AGN activity depends acutely on an assortment of galaxy properties: mass, luminosity, color, morphology, and concentration, among others~\citep{Kauffmann2003b,Best2005,Kewley2006,Ellison2008,Choi2009,Silverman2009}. 

To remedy these interrelated dependencies, termed contamination, we introduce a matching procedure. Comparison of the matched galaxies leads to $\Delta$$D$, or relative change in $D$. $\Delta$$D$ measures how activity, most related to ionization field hardness, of AGN varies between two populations; in our case, those residing in different environments.

\subsection{\texorpdfstring{$\Delta$$D$}{Delta D}}
\label{subsec:DeltaD}
As shown in Eq.~\ref{eq:DeltaD}, $\Delta$$D$ is the difference between the $D$ of a sample galaxy and the median $D$ of all matched control galaxies. Specifying a sample and control set that differ only in one property allows us to directly probe the relationship between that property and $D$. Matching control galaxies to each sample galaxy on other specified properties (e.g., galaxy mass, redshift, and star formation rate) lets us marginalize the contaminating contribution of those properties to reduce the impact of other effects mimicking that due to the environment. Taking the difference in the measure between the sample galaxy and matched control galaxies (aggregated by taking their median) then gives change, or variation, of the measure of activity. The calculation is performed for every sample galaxy, forming a corresponding matched control subset for each from the total control set. Details of each stage of this procedure are given below.

\footnotesize
\begin{equation} 
\Delta D=D_\text{sample}-\textit{median}(D_\text{mat. control 1},\ldots,D_\text{mat. control $n$}) 
\label{eq:DeltaD}
\end{equation}
\normalsize

First, the sample and control sets are established as subsets of the whole dataset. The two must be considered jointly, since $\Delta$$D$ is a measure of the relative difference in AGN activity between the sample and its control. Thus key conditions are emphasized and their effect on the AGN's activity can be isolated. For example, to probe the effect of pair interactions on AGN activity, the sample is set as all pair galaxies and the control as all nonpair galaxies. So, the change going from control to sample is the state of being in a pair. 

Second, a subset of control galaxies is matched to each sample galaxy based on specific galaxy properties lying within a predefined matching tolerance. These properties can correlate with $D$ but do not necessarily reflect AGN activity, especially ionization state. Since they are kept constant between control and sample, their contribution is marginalized. Matching on mass effectively controls for all sensitive galaxy properties~\citep{Ellison2011}, and mass itself shows the strongest trend with AGN fraction~\citep[e.g.,][]{Sabater2013}. Matching on redshift is also done in order to address aperture bias and any dependency of cluster richness on redshift. Controlling for star formation rate is accomplished by matching on the $D_{4000}$ break index, which has lower values corresponding to higher rates of star formation~\citep{Poggianti1997}. 

The matching tolerance for each parameter is based on the strength of its correlation with $D$ and the parameter's inherent range of values. Higher correlation and smaller range necessitate a stricter tolerance, but there is no exact formula. We set matching tolerances of 0.1 dex $M_{\sun}$ for mass, 0.01 for redshift, and 0.1 for $D_{4000}$. 

Third, the median $D$ of the matched control galaxies is subtracted from the $D$ of the sample galaxy to give $\Delta$$D$ for that sample galaxy. Aggregating the controls by taking the median rather than average further reduces the influence of outliers in the control sample. In order to remove outliers created by small control samples, a minimum of 5 matches is required to be included in the results.

The general procedure of comparing a sample galaxy to a set of matched controls, in order to emphasize certain properties while marginalizing others, has been successfully employed earlier~\citep[e.g.,][]{Ellison2013}. In this study, however, $D$ is chosen as the measure of activity for the first time, so the behavior of $\Delta$$D$ must be examined.

It is important to consider the difference between the absolute and relative meanings of $\Delta$$D$, since as a comparative measure, it behaves differently than typical observables. Since the placement of an AGN galaxy on the BPT diagram is a nonlinear function of activity, $D$ is likely a nonlinear measure. Then, the change in $D$ would be nonlinear as well. So, $\Delta$$D$ does not have an absolute meaning to its value (i.e., it depends on position on the BPT diagram) and its scaling is nonlinear. Nonetheless, $\Delta$$D$ always has a relative meaning since $D$ increases monotonically with radiation field hardness and ionization parameter~\citep{LaMassa2012,Kewley2006}. A greater absolute value of $\Delta$$D$ should always indicate more change in radiation field hardness than a lesser value when different samples are compared to the same control.

Physically, if $D$ is a probe of the incident radiation field hardness, then $\Delta$$D$ likely indicates the change in ionization state of the AGN between a specific galaxy and other, similar galaxies. If $D$ is taken as an indication of the balance between AGN and SF activity, matching on $D_{4000}$ focuses on the AGN contribution. $\Delta$$D$ should therefore track changes in the ionization state of the AGN. The ionization state is strongly correlated with the luminosity of the AGN~\citep{Kauffmann2003}, corresponding to accretion rate. Thus, we interpret $\Delta$$D$ as the change in ionization state of the AGN, which is correlated with other indicators of its strength such as its accretion rate.

\begin{figure}[!hbtp]
  \centering
  \leavevmode
  \includegraphics[width={\eps@scaling\columnwidth}]{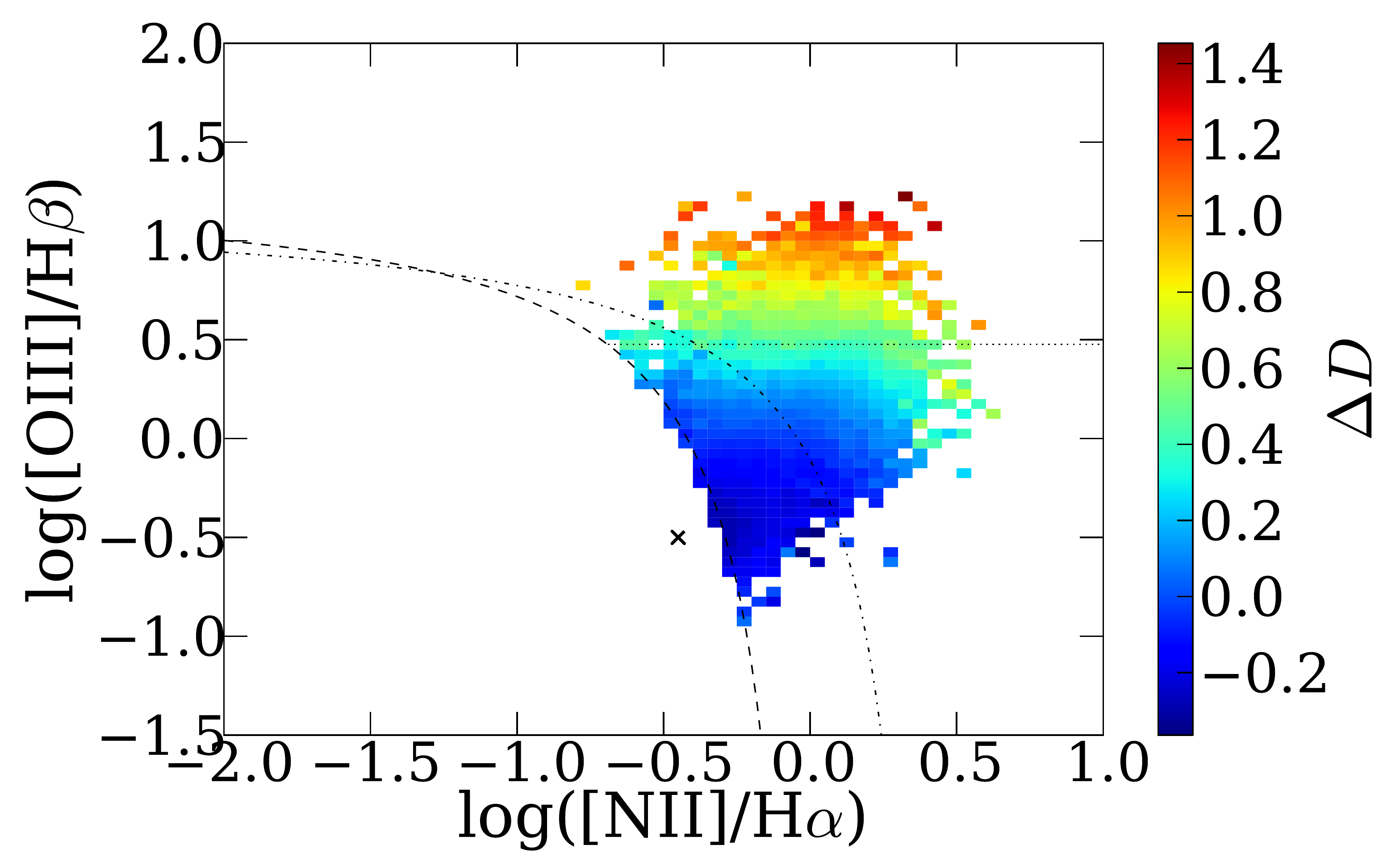}
  \caption{BPT diagram of sample galaxies for the case of matching a random tenth of the dataset with itself. The galaxies are colored by $\Delta$$D$ and averaged into rectangular bins on the [O{\scriptsize III}]/H{\scriptsize $\beta$} and [N{\scriptsize II}]/H{\scriptsize $\alpha$} coordinate axes. The SF locus (black x), \citet{Kauffmann2003} classification curve (dashed black curve), and \citet{Kewley2001} classification curve (dash-dotted black curve) are included as a guide. AGN galaxies above $\text{[O{\scriptsize III}]/H{\scriptsize $\beta$}}=3$ (dotted black line) are traditionally identified as Seyferts while those below are LINERs~\citep[e.g.,][]{Kauffmann2003}. On average, sample galaxies smoothly increase in $\Delta$$D$ going from LINERs to Seyferts (mirroring the increase in ionization field hardness and change in ionization state).}
  \label{fig:DeltaD}
\end{figure}

It is helpful to visually demonstrate the behavior of $\Delta$$D$. We construct a specific example where the sample and control are the same: a random subset with no distinguishing traits, about one tenth the size of the whole dataset. As described above, the matched control galaxies have the same mass as their associated sample galaxy within 0.1 dex, the same redshift within 0.01, and the same $D_{4000}$ break within 0.1. The sample galaxies are plotted on a BPT diagram and colored with the $\Delta$$D$ resulting from the matching procedure (Fig.~\ref{fig:DeltaD}). Averaged bins are displayed due to the large number of galaxies. The $\Delta$$D$ distribution exhibits a smooth vertical gradient of increasing $\Delta$$D$ going from low-ionization states (LINERs) to high-ionization states (Seyferts). This sequence has increasing ionization field hardness and supports $\Delta$$D$'s physical interpretation as a continuous measure of change in ionization state. Models of the BPT diagram support the illustrated trend by suggesting that with increasing hardness of the ionizing radiation field, galaxies move up and to the right on the diagram~\citep{Kewley2013}.

$\Delta$$D$ is an improvement over using $D$ because it reduces contamination from interrelated dependencies, focuses on the impact of specified conditions, and measures change in AGN activity, allowing sensitive searches as described in later sections. The drawbacks are a more complicated procedure and nonstandardized meaning to $\Delta$$D$. The analysis of BPT diagrams above supports our use and interpretation of $\Delta$$D$: a relative measure of change in AGN state. In \S~\ref{subsec:Merging}, using $\Delta$$D$ reproduces a widely accepted result, providing additional validation. For these reasons, we adopt $\Delta$$D$ along with its methodology.  

\section{Results}
\label{sec:Results}

In this section, we use $\Delta$$D$ to perform a series of studies to isolate the effect of different environments on AGN activity. The sample and control are set for each study to isolate the effect of a specific environmental condition. Binning and plotting on a measure of this condition is done to investigate trends as well as asymptotic behavior where sample galaxies approximate control galaxies. In these studies, the binned variable is always distance, either to the companion pair galaxy or to the center of a cluster. The following results were checked to be consistent over the full redshift range and with average used instead of median in the calculation of $\Delta$$D$.

In each figure in this section, the indicated $\Delta$$D$ for each bin is the average $\Delta$$D$ of all sample galaxies in that bin. An offset is added in order to choose the zero value of $\Delta$$D$ as the average over all sample galaxies matched to themselves. Uncertainties for each bin are estimated using bootstrap resampling, since calculating $\Delta$$D$ for each sample galaxy involves taking a median and precise uncertainties are unknown. In general, the uncertainties appear to be overestimated as they tend to exceed the scatter in plots, which could indicate that data in different bins are not completely independent of each other. This correlation may result from difficulties in measuring the binned variable due to, for example, projection effects. In interpreting these results, it is important to remember that linear progression along a binned variable does not necessarily imply linear progression in an environmental process, especially for pair interactions that might involve multiple close approaches of a pair of galaxies prior to an eventual merger. 

\subsection{Pairs Induce Increase in Activity}
\label{subsec:Merging}

\begin{figure}[!hbtp]
  \centering
  \leavevmode
  \includegraphics[width={\eps@scaling\columnwidth}]{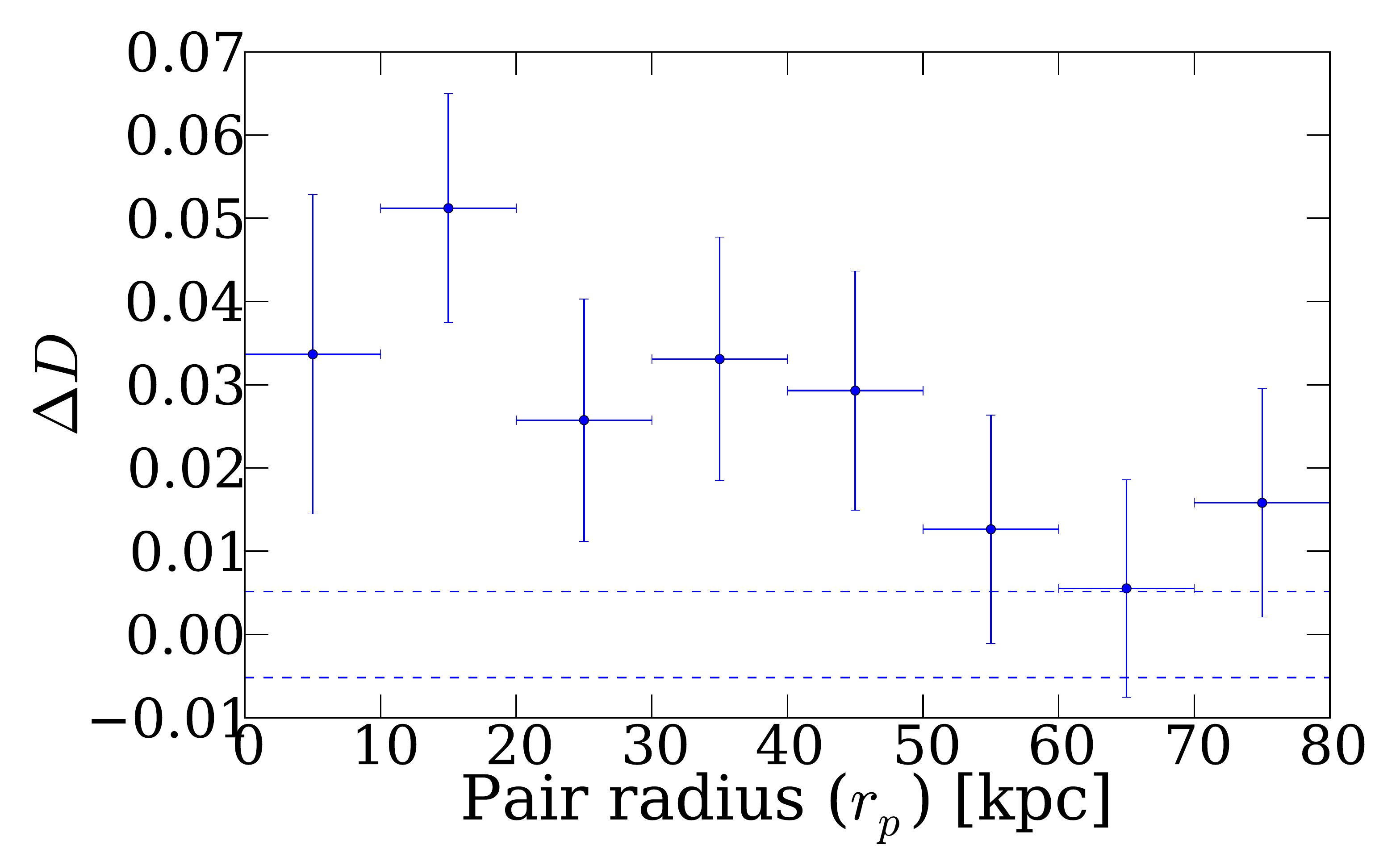}
  \caption{$\Delta$$D$ vs pair radius for pair galaxies compared to nonpair galaxies. Dashed lines indicate the uncertainty in the 0 value. There is increasing activity with decreasing pair separation that begins around $r_p=80$ kpc, where pair galaxies approximate nonpairs. Taken over all radii on the unbinned data, the Pearson correlation coefficient is \sn{-4.6}{-2} with a two-tailed p-value of \sn{9.5}{-3}.}
  \label{fig:Mergers}
\end{figure}

We consider the relationship between close galaxy pairs and AGN activity. Previous studies indicate that AGN activity becomes stronger in close pairs, reporting increases in AGN fraction~\citep{Silverman2011,Ellison2011,Koss2010,Woods2007} and accretion rate~\citep{Ellison2013,Satyapal2014} with decreasing pair separation ($r_p$). Using $\Delta$$D$ by comparing a sample of pair galaxies to a control of nonpair galaxies, we independently produce a similar result, illustrated in Fig.~\ref{fig:Mergers}. Increasingly positive $\Delta$$D$  corresponds to stronger AGN activity with decreasing separation. This trend should approach $\Delta$$D$$=0$ at large separations, where the influence of the galaxy pair is negligible.  We observe an almost complete return to the control value by $r_p=80$ kpc. 

Pairs have been found to have an effect at separations greater than 80 kpc for properties such as metallicity, star formation rate, and color~\citep{Patton2011,Scudder2012}. \citet{Patton2013} find increases in star formation rate out to 150 kpc and \citet{Foreman2009} observe increases in the rate of double quasars out to 100-200 kpc. Our study indicates that substantial changes in ionization state of AGN require smaller separations. Given the uncertainties for $\Delta$$D$ close to 0, however, we can make no determination about whether a modest increase might persist out to 100-200 kpc using our current sample.

\subsection{Clusters' Suppression of Activity}
\label{subsec:Clusters}

We investigate the effect of being in a cluster on AGN activity by comparing a sample of cluster AGN to a control of noncluster AGN, without regard for the local influence of pairs.  Previous studies have produced conflicting results by measuring AGN fraction, reporting increases~\citep{Ruderman2005} as well as decreases~\citep{Pimbblet2013} at small distances from the cluster center. Our analysis is distinct in that we measure continuous changes in ionization state. In addition, galaxy density has often been used to measure the strength of the cluster environment. These types of studies usually suggest that AGN fraction is constant going from the field to the center of the cluster~\citep{Miller2003,Sorrentino2006}. Instead, we consider a member galaxy's location within the cluster, defining cluster radius as the galaxy's distance to luminosity-weighted cluster center ($r_c$) divided by the cluster's $R_{200}$ value~\citep[cf.][for the determination of these two distances]{Yang2005}. 

The influence of cluster size is evaluated by comparing two subsamples: small and large clusters. Cluster size is evaluated two ways. First, we consider two subsamples based on group halo mass: light clusters ($<14.5$ dex $M_{\sun}$) and heavy clusters ($\geq 14.5$ dex $M_{\sun}$). Second, we consider two subsamples based on richness: poor clusters ($<25$ members and $\geq 10$ members to be classified a cluster) and rich clusters ($\geq 25$ members). The same control of noncluster galaxies is always used.

\begin{figure}[!hbtp]
  \centering
  \leavevmode
  \includegraphics[width={\eps@scaling\columnwidth}]{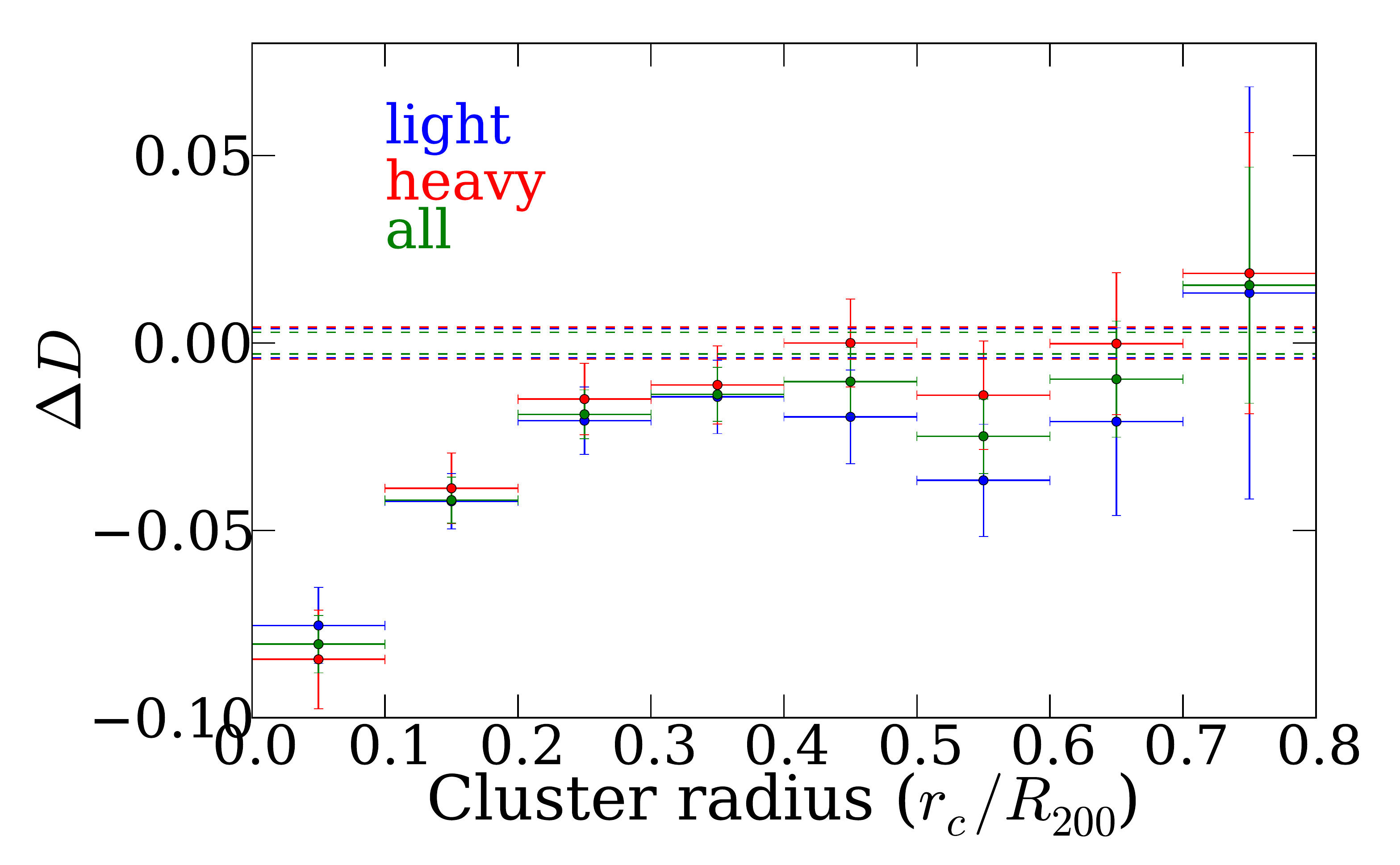}
  \caption{$\Delta$$D$ vs cluster radius for light ($<14.5$ dex $M_{\sun}$), heavy ($\geq 14.5$ dex $M_{\sun}$), and all cluster galaxies compared to noncluster galaxies. Dashed lines indicate the uncertainty in the 0 value. There is, on average, decreasing activity with decreasing cluster radius. At higher cluster radii, where cluster galaxies approximate nonclusters, $\Delta$$D$ becomes asymptotic to 0. The trend has no significant dependence on halo mass. Taken over all radii on the unbinned data, the Pearson correlation coefficient is \sn{5.7}{-2} with a two-tailed p-value of \sn{5.9}{-5} (light), \sn{7.3}{-2} with a two-tailed p-value of \sn{8.8}{-7} (heavy), and \sn{6.7}{-2} with a two-tailed p-value of \sn{6.4}{-11} (all).}
  \label{fig:D_cr_both}
\end{figure}

\begin{figure}[!hbtp]
  \centering
  \leavevmode
  \includegraphics[width={\eps@scaling\columnwidth}]{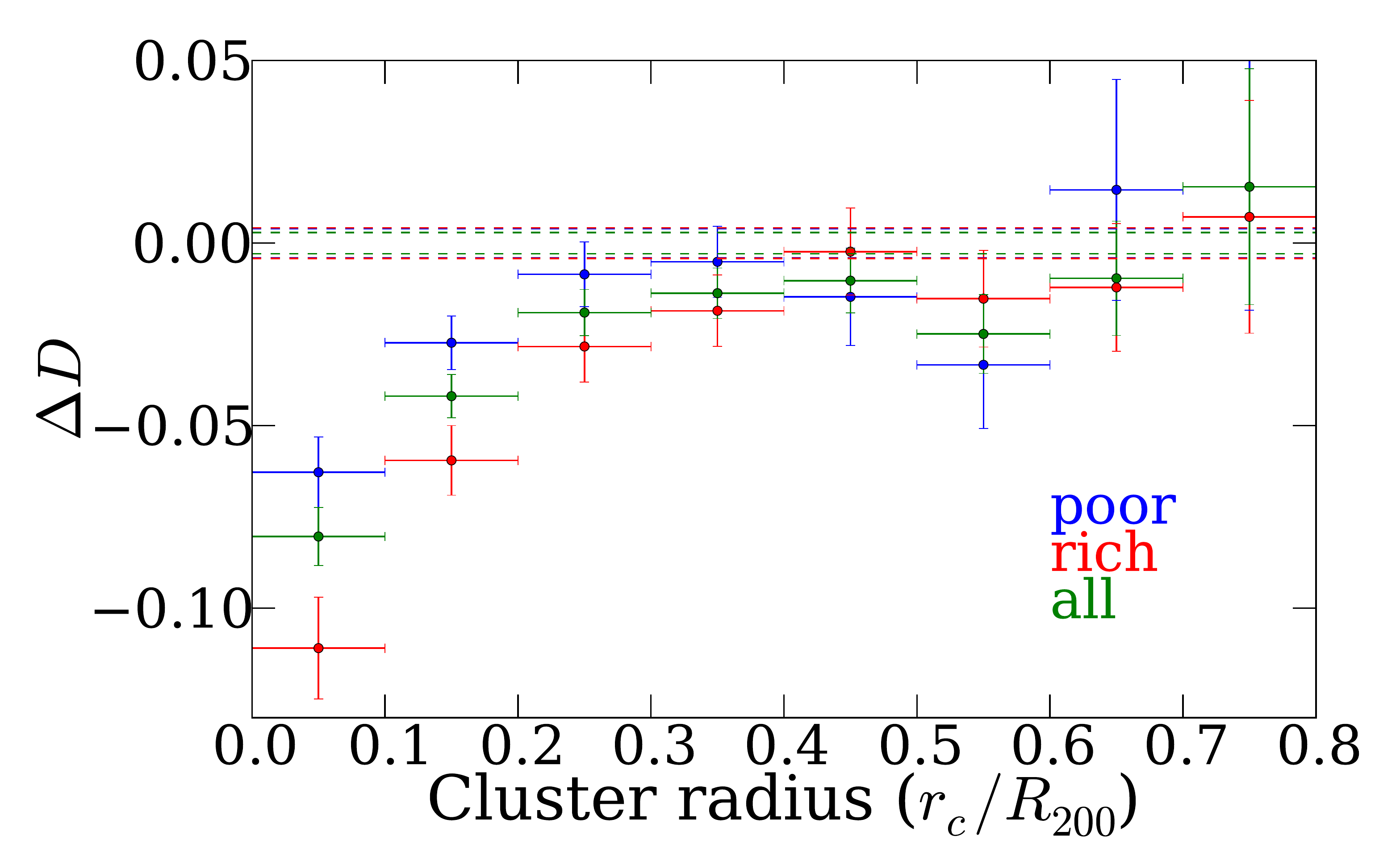}
  \caption{$\Delta$$D$ vs cluster radius for poor ($<25$ members), rich ($\geq25$ members), and all cluster galaxies compared to noncluster galaxies. Dashed lines indicate the uncertainty in the 0 value. There is, on average, decreasing activity with decreasing cluster radius. At higher cluster radii, where cluster galaxies approximate nonclusters, $\Delta$$D$ becomes asymptotic to 0. The decrease is greater for richer clusters. Taken over all radii on the unbinned data, the Pearson correlation coefficient is \sn{5.4}{-2} with a two-tailed p-value of \sn{1.6}{-4} (poor), \sn{9.6}{-2} with a two-tailed p-value of \sn{8.6}{-11} (rich), and \sn{6.7}{-2} with a two-tailed p-value of \sn{6.4}{-11} (all).}
  \label{fig:D_cr_r_both}
\end{figure}

Overall, we find evidence for decreasing ionization strength (negative $\Delta$$D$) at lower cluster radii, most pronounced below $r_c/R_{200}<0.4$. That is, proximity to the center of a cluster suppresses AGN activity. This can seen for cluster samples split by either halo mass (Fig.~\ref{fig:D_cr_both}) or richness (Fig.~\ref{fig:D_cr_r_both}). There does not appear to be a significant change with halo mass (Fig.~\ref{fig:D_cr_both}) except for a dip in the outer region which disappears at smaller radii; this feature is isolated to one bin so probably not real. Richness does (Fig.~\ref{fig:D_cr_r_both}) have a salient effect. The richer the cluster, the greater the decrease in the inner region. Dependence on richness but not halo mass suggests that the effect is based on local interactions rather than global cluster properties. The radii where the decrease occurs do not depend on either mass or richness. Beyond the outer region of decrease ($r_c/R_{200}>0.7$) cluster galaxies approximate the field with no change in ionization state.

\subsection{Distribution of Pairs in Clusters}
\label{subsec:ClusterComposition}

\begin{figure}[!hbtp]
  \centering
  \leavevmode
  \includegraphics[width={\eps@scaling\columnwidth}]{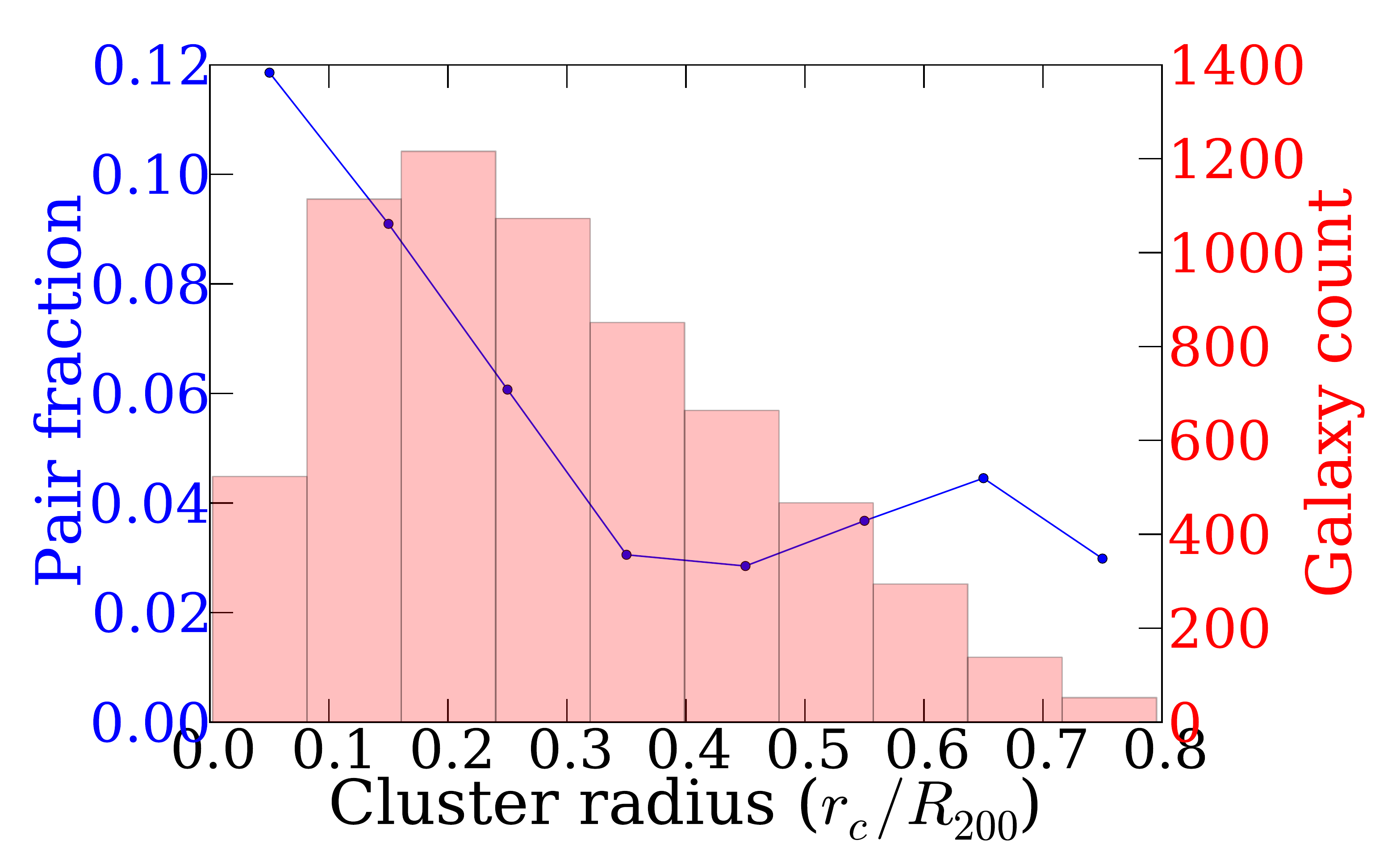}
  \caption{Pair fraction (blue line) and galaxy count (red histogram) in clusters. With decreasing cluster radius, pair fraction increases, except for a dip at $r_c/R_{200}=0.4$. The total number of AGN galaxies also increases until an expected drop to 0 at the center. These figures imply a significant increase in the number of AGN in pairs inside clusters. However, the pair fraction may be confounded by increased projection effects and high speed interactions toward the center of clusters. No vertical error in pair fraction is indicated because the data points are counts per bin, but bins with more galaxies are more reliable.}
  \label{fig:ClusterComposition_Both}
\end{figure}

Having investigated the effects of pair interactions and cluster environment independently above, a more general picture of environmental influence may be formed by studying the two together. The large-scale decreases in AGN activity inside clusters could be explained by small-scale pair-induced increases if pair frequency decreases inside clusters. We examine this possibility by observing how AGN in pairs and AGN galaxies in general are distributed in clusters (Fig.~\ref{fig:ClusterComposition_Both}). An initial examination of the data indicates that AGN are increasingly likely to be in close pairs near the cluster center (see the following paragraph for alternative explanations). There is a dip in pair fraction around $r_c/R_{200}=0.4$, such that the regions that have a relatively high fraction of AGN galaxies in pairs are $r_c/R_{200}<0.3$ and $0.6<r_c/R_{200}<0.7$. These roughly match the regions of decreased activity inside clusters. At large cluster radius, cluster galaxies approximate field galaxies so an AGN is more likely to be in a pair inside clusters than outside. In addition, with decreasing cluster radius the total number of AGN increases until the very center, so the rise in the number of AGN in pairs is significant. 

Before drawing physical conclusions, we caution that the plotted pair fraction suffers from contamination that cannot be removed from the analysis and is hard to estimate. Some of the ``pairs'' might be due to projection effects, so that in reality the galaxies are too physically separated to interact. In addition, some nearby galaxies classified as being in pairs may be passing each other at high speeds, resulting in a flyby with negligible interaction. The influence of these two effects likely increases toward the center of clusters, where there are more galaxies and higher velocities. Therefore, the true fraction of AGN involved in pair interactions likely is over-represented, especially closer to the cluster center.

We might speculate about several physical interpretations. If the number of interacting pairs does not truly increase toward the cluster center, but rather decreases, then pair-induced increases in AGN activity could explain cluster-induced decreases. If it remains constant, then the influence of clusters may be a large-scale environmental one. If the number does in reality increase towards the cluster center, a corresponding pair-driven increase (Fig.~\ref{fig:Mergers}) in activity should be expected, contrary to the decrease indicated by Figs.~\ref{fig:D_cr_both} and \ref{fig:D_cr_r_both}. The correspondence of the regions of increased pair fraction and decreased activity implies entanglement between the effect of pair interactions and the effect of cluster environment.

An increase of galaxies in pairs in clusters seems to contradict the previous results, motivating further study in the following sections. One possibility is that pair interactions act differently on AGN inside clusters. If there they actually decrease AGN activity, then the results presented so far are reconcilable.

\subsection{Pairs in Clusters Induce a Weakened Increase in Activity}
\label{subsec:MergersInClusters}

\begin{figure}[!hbtp]
  \centering
  \leavevmode
  \includegraphics[width={\eps@scaling\columnwidth}]{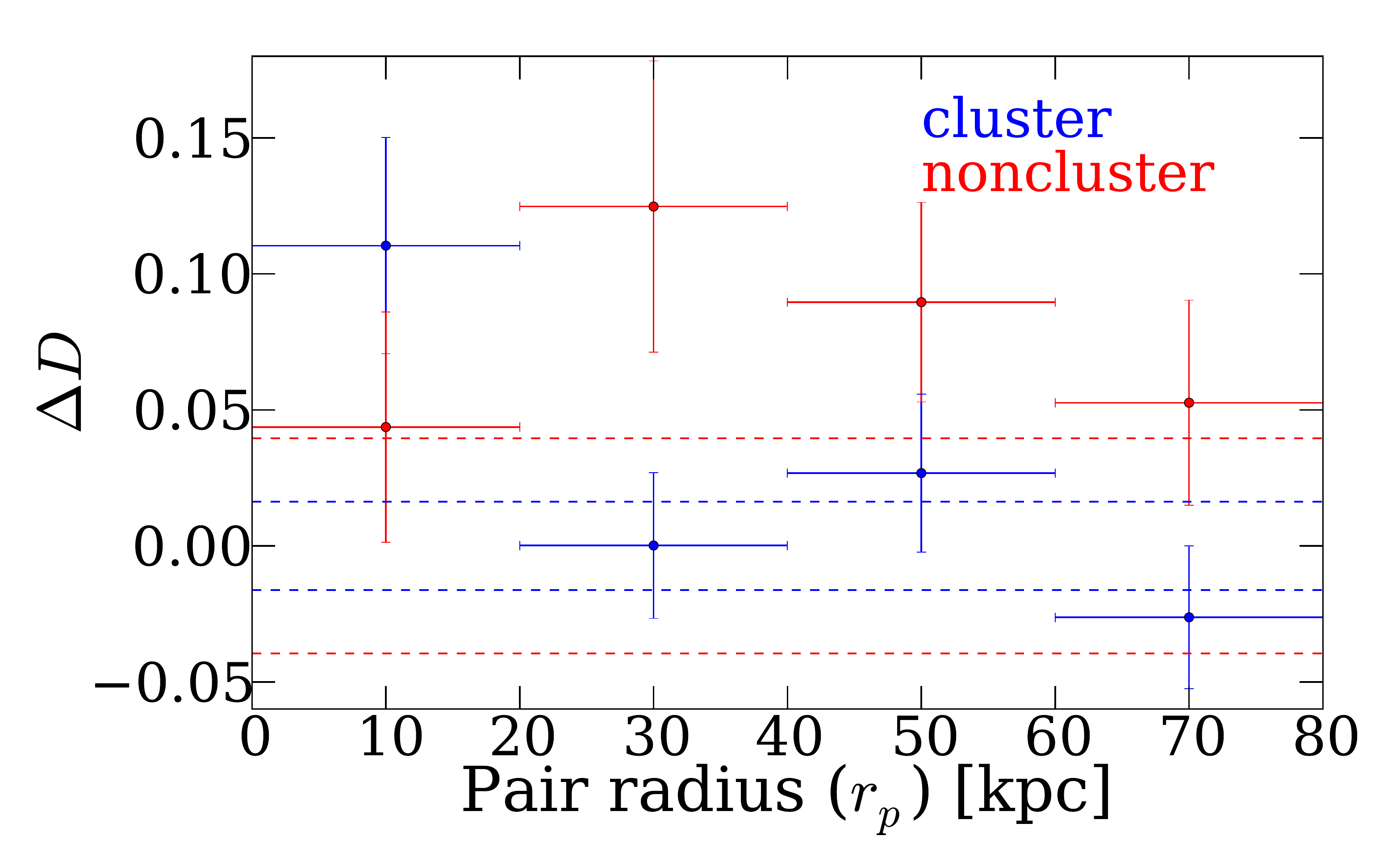}
  \caption{$\Delta$$D$ vs pair radius for pair AGN inside clusters compared to nonpair AGN inside clusters and pair AGN outside clusters compared to nonpair AGN outside clusters. Dashed lines indicate the uncertainty in the 0 value. Pair-induced increases in activity are present in both environments, but are much weaker inside clusters, except for the smallest radii. Taken over all radii on the unbinned data, the Pearson correlation coefficient is \sn{-1.2}{-1} with a two-tailed p-value of \sn{1.5}{-2} (inside clusters) and \sn{-5.0}{-2} with a two-tailed p-value of \sn{4.6}{-1} (outside clusters).}
  \label{fig:D_r_ncrc_plot}
\end{figure}

To understand the effect of cluster environment on close pair interactions, we consider two subsamples of AGN: those inside and outside clusters (Fig.~\ref{fig:D_r_ncrc_plot}). Both show increases in activity, but behave in contrasting ways. The average $\Delta$$D$ inside clusters is \sn{1.6}{-2} while the average $\Delta$$D$ outside clusters is \sn{7.8}{-2}. Thus, the average induced increase in activity due to being in a pair is smaller inside clusters. In addition, the increase inside clusters occurs only at very small pair separations ($r_p<20$ kpc), while the increase starts from at least $r_p=80$ kpc outside clusters. By $r_p=20$ kpc the increase outside clusters drops, but given the uncertainty this drop may not be statistically significant. Taken together, the results of this and the preceding subsections indicate that cluster environment generally decreases AGN activity (most closely, ionization field hardness) and weakens pair-induced increase, although close pair interactions still promote a positive change.

This result does not resolve the apparent contradiction posed at the end of \S~\ref{subsec:ClusterComposition}: Clusters induce decreasing activity toward their centers yet have more AGN involved in pair interactions there, which may induce small increases in activity. However, the two results would be compatible if clusters have an independent large-scale effect that dominates the small-scale (i.e., local), relative influence of pairs. AGN in central pairs may very well have reduced activity compared to AGN outside clusters, and then central AGN not in pairs would have even further, albeit slightly, reduced activity. To test whether this is the case, we continue to explore the combined dynamics of pairs and clusters by conducting a study at the scale of clusters.

\subsection{Cluster-Induced Decrease Dominates Pair-Induced Increase in Activity}
\label{subsec:CombinedMergersAndClusters}

\begin{figure}[!hbtp]
  \centering
  \leavevmode
  \includegraphics[width={\eps@scaling\columnwidth}]{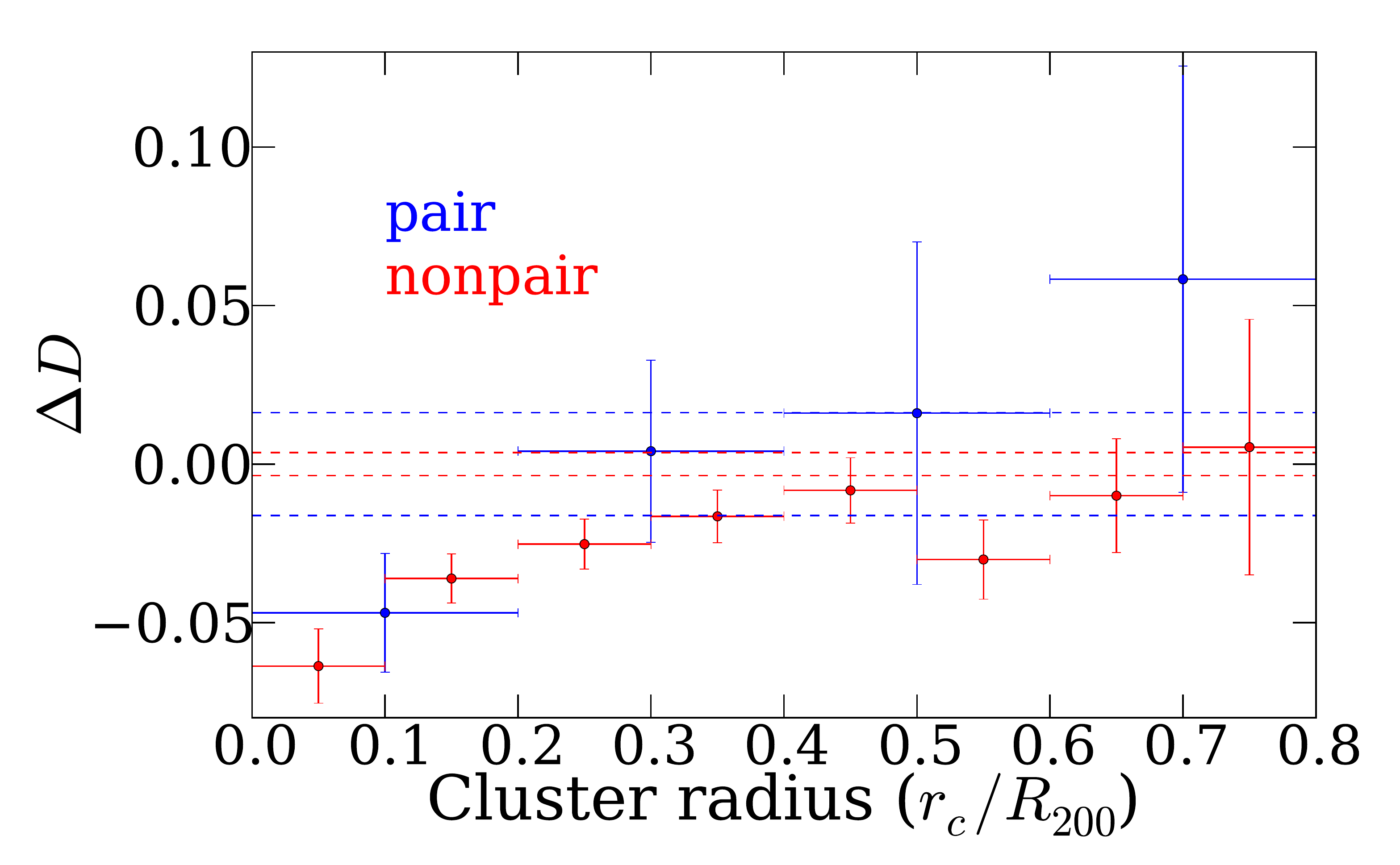}
  \caption{$\Delta$$D$ vs cluster radius for pair AGN and nonpair AGN in clusters compared to noncluster nonpair AGN. Dashed lines indicate the uncertainty in the 0 value. At the edges of clusters, pairs have the typical increase in activity while nonpairs approximate the field (comparable to Fig.~\ref{fig:Mergers}), so the pair-induced effect is most important while the cluster environment is still weak. Near the centers of clusters, there is a decrease in activity for pair AGN that approaches the decrease for nonpair AGN (comparable to Figs.~\ref{fig:D_cr_both} and~\ref{fig:D_cr_r_both}). Thus, the cluster-induced effect becomes dominant over the pair-induced effect. Taken over all radii on the unbinned data, the Pearson correlation coefficient is \sn{1.0}{-1} with a two-tailed p-value of \sn{3.9}{-2} (pair) and \sn{4.5}{-2} with a two-tailed p-value of \sn{5.2}{-4} (nonpair).}
  \label{fig:D_cr_x_nn}
\end{figure}

We now set the control to be all field AGN (i.e., outside clusters, not in pairs). Since this group contains the large majority of all galaxies, the following result (Fig.~\ref{fig:D_cr_x_nn}) should be compared to the studies of general pair-induced increase (Fig.~\ref{fig:Mergers}) and general cluster-induced decrease (Figs.~\ref{fig:D_cr_both} and~\ref{fig:D_cr_r_both}), where the controls are approximately the same. We set the sample as cluster AGN, either in pairs or not. Nonpair AGN in clusters have the average trend of a cluster-induced decrease in activity. The large-scale effect of clusters on AGN in pairs can now be considered in comparison with these non-pair cluster AGN, although the uncertainty is large due to the small number of AGN in pairs in clusters. 

At the outskirts of clusters (high $r_c/R_{200}$), cluster galaxies approximate noncluster galaxies and so the expected pair-induced increase is reproduced. With decreasing cluster radius, the cluster environment becomes increasingly important. Looking at AGN in close pairs, we observe a gradual shift from positive $\Delta$$D$ to negative $\Delta$$D$ that approaches the values for nonpairs. That is, the cluster-induced reduction in activity for pair galaxies approaches that of nonpair galaxies toward the center of the cluster. Cluster galaxies near the outskirts of the cluster appear typical of field galaxies and display the expected independent environmental effects, but closer to the cluster center any expected pair-induced increase in activity is gradually replaced by cluster-induced decrease. The effect on an AGN's ionization state of being in a cluster is dominant over the effect of being in a pair.

\section{Discussion}
\label{sec:Discussion}

We use a new twist on an existing measure, $\Delta$$D$, as an improved indicator of the relationship between AGN activity (most related to ionization state of the AGN) and the environment of the host galaxy to produce three main conclusions: 
\begin{itemize}
\item Pair interactions induce an increase in AGN activity.
\item Cluster environment induces a decrease in AGN activity.
\item Cluster-induced effects dominate pair-induced effects. 
\end{itemize}
These conclusions can be physically explained by gas dynamics. 

The increase in AGN activity due to pair interactions such as merging has been shown in previous studies~\citep[e.g.,][]{Ellison2011,Silverman2011}. Although we study a slightly different aspect of AGN activity by focusing on radiation field hardness, the increase that we observe is likely caused by similar physical processes. Interactions trigger activity by producing tidal torques that mix gas and send it toward the galaxy center \citep{Hernquist1989,Domingue2005,DiMatteo2005}. Extra gas delivery to the central black hole powers accretion and perhaps induces a change to a more efficient ionization state. Outside of the cluster environment, the drop in increased activity at very close separations (Fig.~\ref{fig:D_r_ncrc_plot}) may be attributed to various factors, such as an exhaustion of deliverable gas or obstruction of the delivery process.  This drop may not be statistically significant, however, and the increase may very well continue toward the smallest separations and up to post-mergers, as observed by other studies~\citep{Ellison2013,Satyapal2014}.

In an opposite sense, clusters could induce decreases in AGN activity by limiting gas availability, most likely of cold gas, around the galaxy. This unavailability can be achieved several ways. The amount of gas in general may be lower inside clusters~\citep{Boselli2014}. Ram-pressure stripping~\citep{Fujita2004} in particular can cut a galaxy's available supplies as it falls into the cluster, thus starving the accreting AGN and possibly inducing a less efficient ionization state. Although gas may be sent inward toward the cluster center, it could be rapidly depleted by processes like star formation so that less is available for accretion~\citep{Storchi-Bergmann2006}. Increases in star formation rate are only seen in relatively low density regions~\citep{Ellison2010}, such as on the outskirts of clusters where such depletion could take place. Another possibility is that gas deposited into a cluster experiences shock heating and switches to a hot phase seen in X-ray emission~\citep{Markevitch2007}, in the process strangulating the galaxies in the central regions. Again, there is no gas left for accretion when in close encounters. 

Recent hydrodynamic simulations demonstrate that the fraction of gas-rich galaxies steadily declines from the field to the cluster center~\citep{Cen2014}. A possible auxiliary effect comes from AGN heating though radio jets that keeps the intracluster medium from condensing onto galaxies. This phenomenon occurs for low-luminosity AGN and so it may reduce the chance that galaxies in the inner cluster region have further AGN activity.

Our results that decreases in activity in clusters correspond to increases in pair fraction and that the decreases are greater in richer clusters suggest that the limited available gas is especially exhausted when there are multiple galaxies near each other using up the resources. The insensitivity of AGN to the host group's mass has previously been noted~\citep{Padilla2010}. 

\citet{Li2006} and \citet{Silverman2014} find a higher prevalence of AGN in the cluster's central galaxy. The cluster center in our study is not the same, however, and the effect may be washed out by the satellite population. Still, \citet{Karouzos2014} provide support that clusters may be conducive to AGN by finding a high-density environment preferential to some AGN. A density study by \citet{Sabater2013} reports the same effect for radio AGN but the opposite for optical AGN, which are studied here. We find that the activity of AGN decreases in the center of clusters. Our result is compatible with a higher frequency of AGN if they are in a less efficient ionization state, due to a reduced gas supply.

The final conclusion that cluster-induced effects must be dominant over pair-induced effects follows naturally. In this picture of gas driving accretion and AGN ionization state, pairs act (positively) on delivery and clusters act (negatively) on availability of gas. Availability mechanisms override delivery mechanisms since the availability of gas is a prerequisite for its delivery. Therefore, cluster-induced decrease in activity dominates pair-induced increase. 

In clusters, pair interactions may still provide some increases by acting on remnant gas at very close separations, but in the big picture their contribution is minor, offsetting the cluster-induced decrease only slightly. Further investigation is necessary to provide more significant confirmation. Nonetheless, \citet{Sabater2013} conclude the same for AGN fractions with a similar picture of the gas dynamics. They also find that the effect of pair interactions decreases from Seyferts to LINERs to passive galaxies. Then, the weakened effect of pair interactions that we see inside clusters is linked to the higher prevalence of low-ionization AGN states.

Our results and interpretation of $\Delta$$D$ suggest that environment can induce changes in ionization state of AGN or preferentially produce a certain ionization state. The transition between Seyfert and LINER has been hypothesized~\citep{Ho2005} to be analogous to the transition between high-state and low-state black hole accretion in X-ray binary systems~\citep{Nowak1995}. \citet{Kewley2006} find that LINERs and Seyferts form a continuous sequence with ionization conditions that match the X-ray binary model. In addition, they suggest, following \citet{Ho2003}, that the nuclear regions of Seyferts are more gas rich than those of LINERs. These observations support our model of environment modulating gas dynamics that preferentially induce different ionization states of AGN.

We note that all the correlation coefficients in the studies presented here range from around 0.01 to 0.10. Thus, all the environmental effects we describe do not occur in every individual case, but with a random distribution such that over a large sample there arises a small but statistically significant average tendency. It is unclear whether this is due to the stochastic nature of the effects of environment or uncertainties in determining the environment of the AGN galaxy.

\section{Concluding Remarks}
In summary, we find that the activity of the nuclei of active galaxies, presumably tied to the ionization state, varies with intergalactic environment by increasing in pairs and decreasing in clusters, with cluster environment having a dominant effect over pair interactions. We also propose gas dynamics as possible underlying physics, where the activity of the AGN is determined by the amount of gas it can accrete.

The influence of environment on AGN ionization state and activity in general holds much potential for further study. Measuring AGN activity as a function of tidal force or gravitational force, rather than radius, may elucidate which physical dynamics are most important. Analyzing specific subsamples of galaxies will indicate the selectivity of the environmental effect. For example, pair-induced increases may be most dramatic in a pair of a certain mass ratio. Further, the measure of activity can be set to track accretion rate or something other than AGN activity, such as star formation. In addition, studies equivalent to this one can be conducted on focused observational datasets where parameters like cluster size and pair separation are known with a much higher degree of certainty, in order to confirm the results presented here or make them more precise. This work is enhanced by using a continuous measure of AGN activity rather than a Boolean determination via BPT classification, and future studies can benefit from similar analysis.

\acknowledgements 
We thank the anonymous referee for constructive comments. We also thank Surhud More, Steve Bickerton, Claire Lackner, Brian Feldstein, Josh Speagle, and Lisa Kewley for helpful discussions. In addition, ETK is particularly grateful to Richard Ellis for guidance throughout the research project and during the writing process. ETK warmly thanks CLS and JDS for their invaluable mentoring, as well as SLE, JTM, and DRP for their support. Furthermore, the Kavli Institute for the Physics and Mathematics of the Universe and the California Institute of Technology with its Summer Undergraduate Research Fellowships (SURF) program provided key financial and organizational assistance.

\bibliography{ms}

\end{document}